\documentstyle[prl,aps,multicol,epsf]{revtex}
\tighten
\input epsf
\begin{document}
\draft
\title{Sub-gap conductance  in ferromagnetic-superconducting mesoscopic 
structures}
\author{M. Leadbeater$^{(1)}$, C.J. Lambert$^{(3)}$,  R.Raimondi$^{(1,2)}$,
and A.F. Volkov$^{(2,3,4)}$}
\address{ 
$^{(1)}$Dip. di Fisica, Universit\`a di Roma3,
\& $^{(2)}$Istituto di Fisica della Materia,
	Via della Vasca Navale 84, 00146 Roma, Italy\\
 $^{(3)}$School of Physics and Chemistry, Lancaster University, 
	 , Lancaster LA1 4YB, UK\\
	$^{(4)}$ Institute of Radioengineering
and Electronics of the Russian Academy of sciences, Mokhovaya 11, Moscow 103907,
Russia$^{\ast}$
 }
\date{\today}
\maketitle
\begin{abstract}
We study the sub-gap conductance of a ferromagnetic mesoscopic region
attached to a ferromagnetic and a superconducting electrode by means
of tunnel junctions. In the absence of the exchange field,
 the ratio $r= \gamma / \epsilon_T$
of the two tunnel junction resistances determines the behaviour of 
the sub-gap conductance
which possesses a zero-bias peak
for $r\gg 1$ and for $r\ll 1$ a peak at finite voltage. 
We show that
the inclusion of the exchange field leads to a peak splitting for
$r\ll 1$, while it shifts the zero-bias anomaly to finite voltages for 
$r\gg 1$.  
\end{abstract} 

\newcommand{\bdg}{Bogoliubov-de Gennes }
\newcommand{\eqn}{\begin{equation}}
\newcommand{\eqnend}{\end{equation}}
\newcommand{\mb}[1]{ {\mbox{\boldmath{$#1$}}}  }

\begin{multicols}{2}
In the last few years there has been a great deal of activity in 
the study of transport through hybrid superconducting mesoscopic 
structures (S/N)
\cite{lambert98}. Amongst the many interesting findings 
it is worth mentioning the observation of a sub-gap conductance
\cite{kastalsky91} in S/N structures, long-range
\cite{courtois96} and anomalous
 proximity effects \cite{petrashov91} in N-S systems and giant 
oscillations in the electrical 
conductance\cite{petrashov93,pothier94} 
and thermopower\cite{chandrasekhar}
in S/N structures containing 
superconducting and/or normal loops.
 
In parallel with these discoveries,  
ferromagnetic-normal hybrid systems (F/N) 
have stimulated
 the curiosity of the scientific community\cite{fn1} and more recently
combined structures involving both
ferromagnetic and superconducting materials like F/S or F/N/S have been
studied both theoretically\cite{jong,minnesota}  and
experimentally\cite{pannetier}. 
We note here that, in general, the introduction of the exchange field 
leads to a decrease of the Andreev scattering due to
the change in the ratio of the two spin populations. For our case, where the
exchange energy is of order the superconducting energy gap, the density of
states is almost constant and this reduction effect will be small.

In this paper we study how the presence of a finite exchange field affects 
well-known coherent phenomena in disordered N(F)/S systems.
We shall focus attention upon the F/S structure shown in Fig.1, calculating its
electrical conductance 
in the diffusive regime by means of both analytical and numerical 
techniques. The former is based on the use of the quasi-classical Green's
function approach, whereas the latter solves the Bogoliubov de Gennes
equations on a tight-binding lattice. Both of these methods have been used
extensively in the study of N/S systems \cite{lambert98}.
For our structure, we recall  that in the ballistic  limit for a simple NIS system
the conductance is suppressed at voltages below the superconducting energy gap
and the  conductance-voltage curve is featureless. A sub-gap anomaly
in the conductance appears only in the presence of two barriers.  
In the diffusive regime the sub-gap conductance exhibits a peak 
below the superconducting energy gap $\Delta$ even in the absence of a second
barrier, because the disorder itself provides the mechanism for multiple scattering.  
The  position of this peak depends on the particular geometry
and relevant parameters
\cite{volkov93}.
 In the absence of the exchange field, the ratio $r= \gamma / \epsilon_T$
of the two tunnel junction resistances
controls different regimes for the sub-gap conductance: 
 a zero-bias peak is present
for $r\gg 1$ and a finite bias peak for $r\ll 1$. Here $\gamma$ and $\epsilon_T$
are proportional to the conductances of the N'/N and N/S interfaces
respectively.  In what follows we show that
the inclusion of the exchange field (we denote the exchange energy by $h$) 
leads to a peak splitting for
$r\ll 1$ for small $h$ (weak ferromagnetism)
while the zero-bias anomaly is shifted to a finite voltage $eV = h$ for 
$r\gg 1$.

 To numerically model  the structure of Fig.\ref{fig1} we consider 
a dirty normal or 
ferromagnetic region (denoted $N$ or $F$) 
connected to external reservoirs by a clean
semi-infinite normal or ferromagnetic lead (denoted $N'$ or $F'$) 
and a clean superconductor (denoted by $S$).
For our numerical calculation we use 
 a tight-binding version of the \bdg Hamiltonian: 
\eqn
\pmatrix{ H_0 - \sigma h & \sigma\Delta \cr \sigma\Delta^*  & 
-H_0 - \sigma h}\pmatrix{u_\sigma \cr v_{-\sigma} }\, =\, 
\epsilon\pmatrix{u_\sigma \cr v_{-\sigma} }
\label{bdgeq}
\eqnend
where $\sigma =1(-1)$ for spin-$\uparrow$($\downarrow$). In this equation 
$H_0=\sum_{i}|i\rangle \epsilon _{i}\langle i|-t\sum_{\langle 
ij\rangle }|i\rangle \langle j|$ is the standard single--particle Anderson model
with $\langle ij\rangle$ denoting pairs of nearest neighbour sites,
$h=\sum_{i}|i\rangle h_{i}\langle i|$ models the exchange energy 
and $\Delta =\sum_{i}|i\rangle 
\Delta_{i}\langle i|$ is the superconducting order parameter.
The width of the whole structure is $W$ and the length of the disordered
ferromagnet is $d$ (in units of the lattice
constant $a$). The off-diagonal matrix elements $t$, which
determine the width of the energy band, are equal to $1$ throughout
the system except at the $F$/$S$ and $F'$/$F$ interfaces where the value of
$t=t_{FS(F'F)}$ is chosen to model a barrier.
Such barriers may be due  to either the presence of an  insulating layer or to a mismatch 
of the electronic parameters of the adjacent materials. 
Within the
ferromagnetic lead and the superconducting region, the diagonal
matrix elements are $\epsilon _{i}=\epsilon _{0}$ (with $\epsilon
_{0}= 1$, which keeps the Fermi level away from the van Hove
singularity in the band centre). For simulating bulk disorder in
the ferromagnet, the $\epsilon _{i}$ in the scattering region are
taken uniformly from the interval $-U/2<\epsilon _{i}-\epsilon
_{0}<U/2$ where $U$ is the disorder strength.  A numerical decimation
technique \cite{Lambertetc} is used to compute the Greens
function for each realization of disorder. From this  one obtains the
Andreev reflection coefficients $R_{a}^{-\sigma\sigma}$  and
the conductance $G\,=\, R_{a}^{\downarrow \uparrow}+R_{a}^{\uparrow
\downarrow}$\cite{Lambert91}, 
 in units of ${\frac{ e^{2}}{\hbar\pi}}$.

Our analytical treatment is based on the transport theory
described  in 
\cite{volkov93}. There equations for the  quasi-classical Green's functions
were employed in the diffusive regime with the corresponding
boundary conditions at the interfaces
\cite{zaitsev84}.
We begin by recalling the main results  for non ferromagnetic
structures. The normalised conductance $S(V)=(dI/dV)/G_n$ (where $G_n$ is the
conductance of the isolated normal wire) of the
system $N'/N/S$ (exchange energy $h$ is
zero) is equal to (see \cite{volkov93})
\begin{equation}
S(V)=\int_0^{\infty} d\epsilon {F_{\nu}'}{ \left[ m^{-1} (\epsilon)+
{1\over {\nu_d {(\epsilon )}}}\left({{\epsilon_T }\over {\gamma }}\right)
{{\epsilon_d }\over {\epsilon_T}} +{{\epsilon_d }\over 
{M(\epsilon )\epsilon_T }}\right]}^{-1}.
\label{1}
\end{equation}
Here $F_{\nu}'=(\beta /2)[\cosh^{-2}{(\epsilon +eV)\beta}
+ \cosh^{-2}{(\epsilon -eV)\beta}]$ is the derivative of the 
distribution function,
$V$ is the voltage difference between reservoirs,
$\beta =1/2T$,
$\nu_d (\epsilon )=Re (({{\epsilon +i\gamma})/ 
{\sqrt{(\epsilon +i\gamma)^2-\epsilon_T^2}}})$ is the density-of-states
in the central $N$ region, $\gamma ={{\rho D}/{2R_{N\Box}d}}$,
 $\epsilon_T ={{\rho D}/ {2R_{S\Box}d}}$, $\rho$ and $D$ are the
 specific resistivity and diffusion constant in the middle region,
 $d$ is the length of the middle region,
 $R_{N,S\Box}$ are the $N'/N$ or $N/S$ resistances  per unit area in the normal
 state and $\epsilon_d =D/d^2$ is the Thouless energy. We assume here
 that the length $d$ is small enough and the condition
 $\epsilon_d \gg T$ is fulfilled (short contact).
 The coefficient 
 $m(\epsilon)$ determines the conductance of the normal wire and
 is given by
 \begin{equation}
 m^{-1} (\epsilon)={1\over 2}\left[1+{\vert\epsilon + i\gamma\vert^2+
 \epsilon_T^2\over \vert (\epsilon+i\gamma)^2-\epsilon_T^2\vert}\right].
 \end{equation}
 The coefficient $M(\epsilon )$ in eq.(\ref{1}) 
determines the energy dependence of the $N/S$ interface conductance. 
For example, from   eq.(107) of Ref.\cite{lambert98}, we have
\begin{eqnarray}
  M(\epsilon )= Re{{\epsilon +i \Gamma}\over {\sqrt{(\epsilon +i\Gamma)^2-\Delta^2}}}
  Re{{\epsilon +i\gamma}\over {\sqrt{(\epsilon +i\gamma)^2-\epsilon_T^2}}}&\cr
  +Re{{\Delta }\over {\sqrt{\Delta^2-(\epsilon +i\Gamma)^2}}}
  Re{{\epsilon_T}\over {\sqrt{\epsilon_T^2-(\epsilon +i\gamma)^2}}}.&
  \label{1b}
  \end{eqnarray}
  Here $\Delta$ and  $\Gamma$ are the gap and the damping rate in the 
  superconductor,
  respectively.  
  The first term in eq.(\ref{1b}) is due to the quasi-particle contribution
  (at zero temperature it is not zero only if $eV > \Delta$).
  The second term in $M(\epsilon )$ gives the so-called interference current in
  SIS Josephson junctions and leads to a sub-gap conductance in SIN contacts.

 The three terms in square brackets in eq.(\ref{1}) 
 represent the normalized resistances: the first term is the resistance
of the central $N$ region, the second is the resistance of 
the $N'/N$ interface and the third that of the $N/S$ interface.
 The quantity $\epsilon_T$ is
a pseudo-gap induced  in the middle region by the proximity effect.
One can see from the expression for $\nu_{d}$ that in the 
case of a small $\gamma$ 
(high resistance of the $N'/N$ interface ) $\nu_d$ 
has a form typical for a superconductor with 
the energy gap $\epsilon_T$.
If $\gamma$ is
large compared with $\epsilon_T$,   the singularity in the density of the states 
in the $N$ film disappears due to the strong damping caused by the $N'$ 
electrode.
 In Fig.2 we plot the dependence of the normalized conductance 
 $S(V)$ on the applied voltage $V$  at zero
 temperature assuming that the resistance of the $N$ region is small 
 ($\epsilon_d /\epsilon_T=10$).   
 One can see that in the case of large $\gamma$, the main contribution  
 to $S(V)$ is due to
 the last term in brackets ($N/S$ interface resistance)
  and the 
 sub-gap conductance has a zero-bias peak. In this case there is a gapless
 state in the central $N$ region. 
When $\gamma$ is decreased the main contribution to
 $S(V)$ is
 given by the second term in brackets; the sub-gap appears in the central $N$
 region and the peak is shifted to $\epsilon_T$: 
 the conductance is just that expected for tunneling into a
 superconductor with a energy gap $\epsilon_T$. 
Similar behaviour of the sub-gap
conductance was discussed  in Ref.\cite{volkov94}.
 
 The numerical results in the absence of an exchange field are shown in 
 Fig.\ref{fig3}. To produce this curve we chose $W=d=20$ sites. 
 The barrier at the
 $N/S$ interface was chosen by setting $t_{NS}=0.4$. The different curves
 correspond to $t_{N'N}=0.0,\, 0.1,\, \ldots ,\, 1.0$. The result was obtained
 by averaging over 500 realisations of the random potential with $U=2$ (elastic
 mean free path is 8.5). 
 One can clearly identify
 the transition from finite to zero-bias peak. One should note that for
 these parameters we cannot reach the regime of $\gamma\gg\epsilon_T$ even in
 the absence of the barrier at the $N'/N$ interface 
 and so we do not obtain the Lorenzian
 line shape expected in that limit. This can easily be obtained, however, by
 using a smaller value for $t_{NS}$.
 
 Consider now the case when the central region is a ferromagnet. 
 We describe the
 ferromagnet by adding to the Hamiltonian an exchange term
 $H_{exc}=h\sum_k
 (c^{\dagger}_{k,\uparrow}c_{k,\uparrow}-c^{\dagger}_{k,\downarrow}
 c_{k,\downarrow})$.
 One can write the quasi-classical equations for the
 Green's functions taking into account this term.
 One can easily show \cite{bulaeski} that the energy should be replaced by
 $(\epsilon -h)$ in all terms corresponding to the ferromagnetic regions. 
Therefore eq.(\ref{1}) remains unchanged if we replace $\epsilon$
 by $(\epsilon -h)$ in all such terms (the terms corresponding to the condensate
 functions in the superconductor are unchanged as we do not consider an exchange
 field in these regions). 
Alternatively one can obtain the same expressions for the $F'/F$
 and $F/S$ interface resistances as in eq.(\ref{1}) using the standard 
 tunnel Hamiltonian 
 $H_T=
 U_T\sum_{k,q,\alpha}(c^{\dagger}_{k,\alpha}a_{q,\alpha}
 +a^{\dagger}_{q,\alpha}c_{k,\alpha})$, 
where $a_{q,\alpha}$  are the destruction operators of the reservoirs.
We find the condensate functions  $F^{R(A)}_F$ induced in the ferromagnet  
due to the proximity
effect
\begin{equation}
F^{R(A)}_F ={{\epsilon_T}\over {\sqrt{((\epsilon-h) 
\pm i\gamma)^2-\epsilon_T^2}}}
\label{4}
\end{equation}
where $\epsilon_T =N_S|U_{F/S}|^2$ and  $\gamma
=N_{F'}|U_{F'/F}|^2$. $ N_{S,F'}$ are the density of states in the
$S$ and $F'$ reservoirs in the normal state. Using eq.(\ref{4}) we
come again to the expressions  for the resistances of the $F/S$ and $F'/F$
interfaces (the third and second terms in eq.(\ref{1})).
In Fig. \ref{fig4} we present the voltage dependence of the
conductance $S(V)$ for $h=0.5\epsilon_T$ and for $h=5\epsilon_T$. 
In the former case for all the curves with $\gamma >0.6\epsilon_T$,  the
tunnelling rate due to the ferromagnetic lead is greater than the
exchange energy, and  no peak splitting appears. When $\gamma$
becomes less than the exchange energy, a splitting of the peak
starts to appear. In Fig.\ref{fig4}b,  due to a larger exchange
energy, the zero-bias peak is shifted to finite bias $eV=h$ and for
 small $\gamma$ the splitting of the two peaks is now
 $2\epsilon_T$ reflecting the shift to finite bias of the
 induced gap in the density of states (the gap is now that for one 
 spin--species only). 
 
  To compare with numerical results, we repeat the calculation used to
 obtain Fig.\ref{fig3} but this time there is a finite exchange energy in
 the normal lead and central region. The results are shown in Fig.\ref{fig5} 
 where $h=0.025$. We again see that we have good qualitative
 agreement with the whole curve being shifted to finite bias.

Finally we show that this agreement can be made almost exact. In Fig.\ref{fig6}
we show results for the case $\gamma\gg\epsilon_T$. In this case
equation (\ref{1}) can be approximated by (for $\Delta>\epsilon$ with
$\Gamma=0$)
\begin{equation}
S(V)=\int_0^{\infty} d\epsilon
{F_{\nu}'}\left[{\Delta\over\sqrt{\Delta^2-\epsilon^2}}
{\epsilon_T\over\epsilon_d}{\epsilon_T\gamma\over(\epsilon-h)^2 +
\gamma^2}\right].
\label{FIS}
\end{equation}
We see that the sub-gap conductance is just a Lorenzian: at $T=h=0$ for
$\Delta\gg\epsilon$ equation (\ref{FIS}) is
simply $S(V)=S(0)\gamma^2/((eV)^2+\gamma^2)$. In Fig.\ref{fig6}a
we show the sub-gap conductance obtained numerically with $t_{F'F}=1.0$ 
and $t_{FS}=0.2$ averaged over 500 realisations. 
Each curve corresponds to various
values of $h=0.00,\, 0.01,\,\ldots 0.19$. The value of the gap was 
$\Delta=0.17$.
Each curve has been translated vertically by an amount $0.15$ for clarity. The
curve for $h=0$ can be fitted well by a Lorentzian plus a constant $C$ to take
into account the fact that in the numerical result we cannot neglect completely 
the other terms in equation (\ref{1}). From this we obtain a value
of $\gamma\approx 0.015$ and $S(0)=0.23$ and the constant was $C=0.075$. Putting
$S(0)$ and $\gamma$ into the above formula at $T=0$ 
for the same values of $h$ and $\Delta$ 
used in the numerics gives us Fig.\ref{fig6}b.

In conclusion, we have studied the sub-gap conductance of a F'/F/S structure
for varying ratio of the interface resistances. The presence 
of an exchange field yields: i) a shift to finite bias of the zero-bias peak
when the F/S resistance dominates; ii) a splitting of the finite-bias peak
when the two resistances are comparable. 

R.R. and A.F.V. acknowledge partial financial support 
of INFM under the PRA-project
"Quantum Transport in Mesoscopic Devices" and EU TMR programme
(Contract no. FMRX-CT 960042). 
A.F.V. thanks the EPSRC (England) and the Russian Fund for Fundamental
Research for partial financial support.
M.L. thanks EU TMR programme (Contract no. ERBFMBICT972832).

\newcommand{\figwidth}{7.0truecm}

{\centerline{\epsfxsize=\figwidth\epsfbox{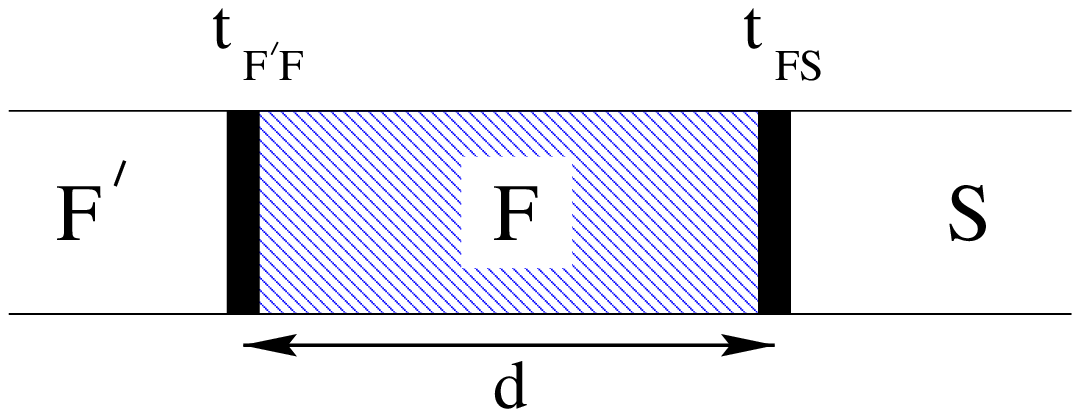}}}
\begin{figure}
\narrowtext
\caption{The system under consideration. }
\label{fig1}
\end{figure}

{\centerline{\epsfxsize=\figwidth\epsfbox{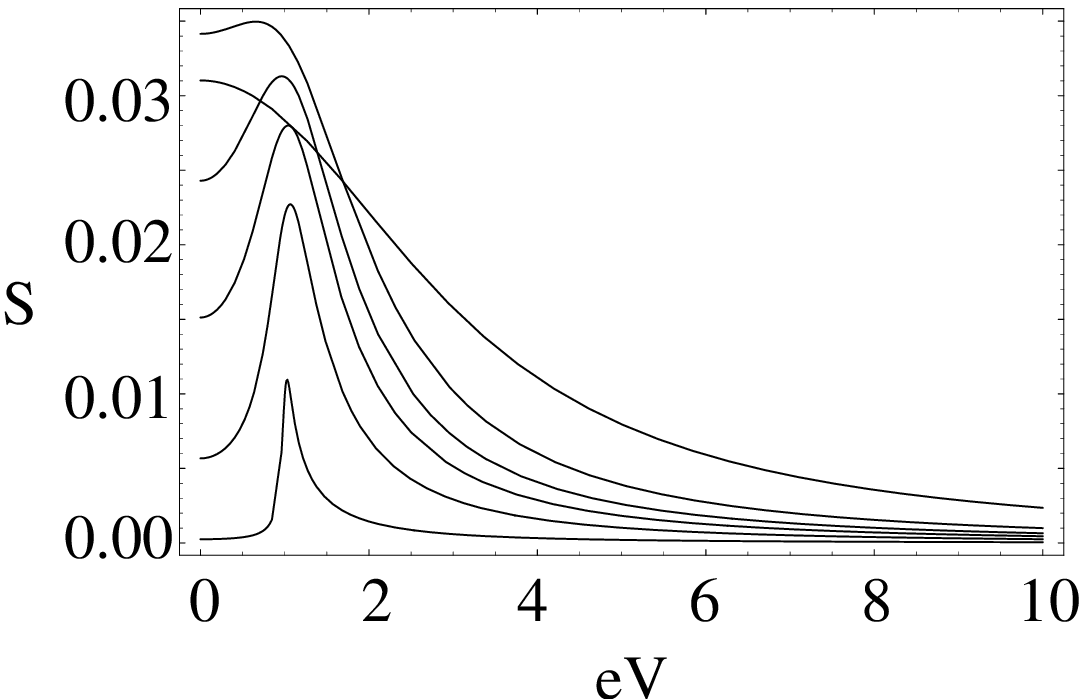}}}
\begin{figure}
\caption{The normalized conductance $S=(dI/dV)/G_n$ as
obtained from quasiclassical theory,  for 
varying  $\gamma $, for $\gamma/\epsilon_T=0.05$ (lower most curve
at large $V$), 0.25, 0.45, 0.65, 1.0 and 2.5 (the uppermost curve
at large $V$). The applied voltage is in units of the
$\epsilon_T$. Here the exchange energy $h=0$.} \label{fig2}
\end{figure}

{\centerline{\epsfxsize=\figwidth\epsfbox{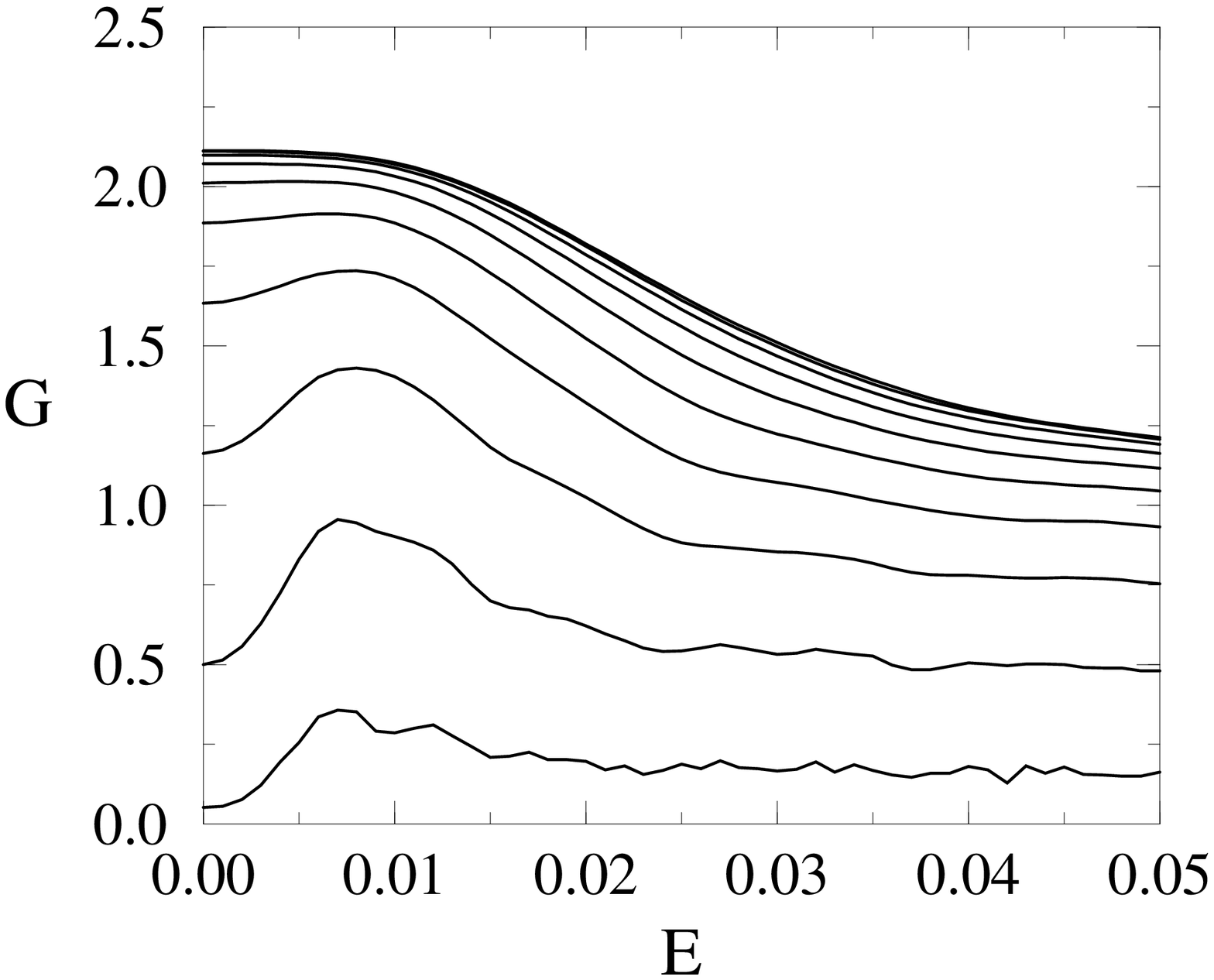}}}
\begin{figure}
\caption{The conductance $G$ as
obtained from numerical simulations,  for 
varying  $\gamma$, from $\gamma>\epsilon_T$ uppermost curve to
$\gamma<\epsilon_T$ lowermost curve. Here the exchange energy $h=0$.}
\label{fig3}
\end{figure}

{\centerline{\epsfxsize=\figwidth\epsfbox{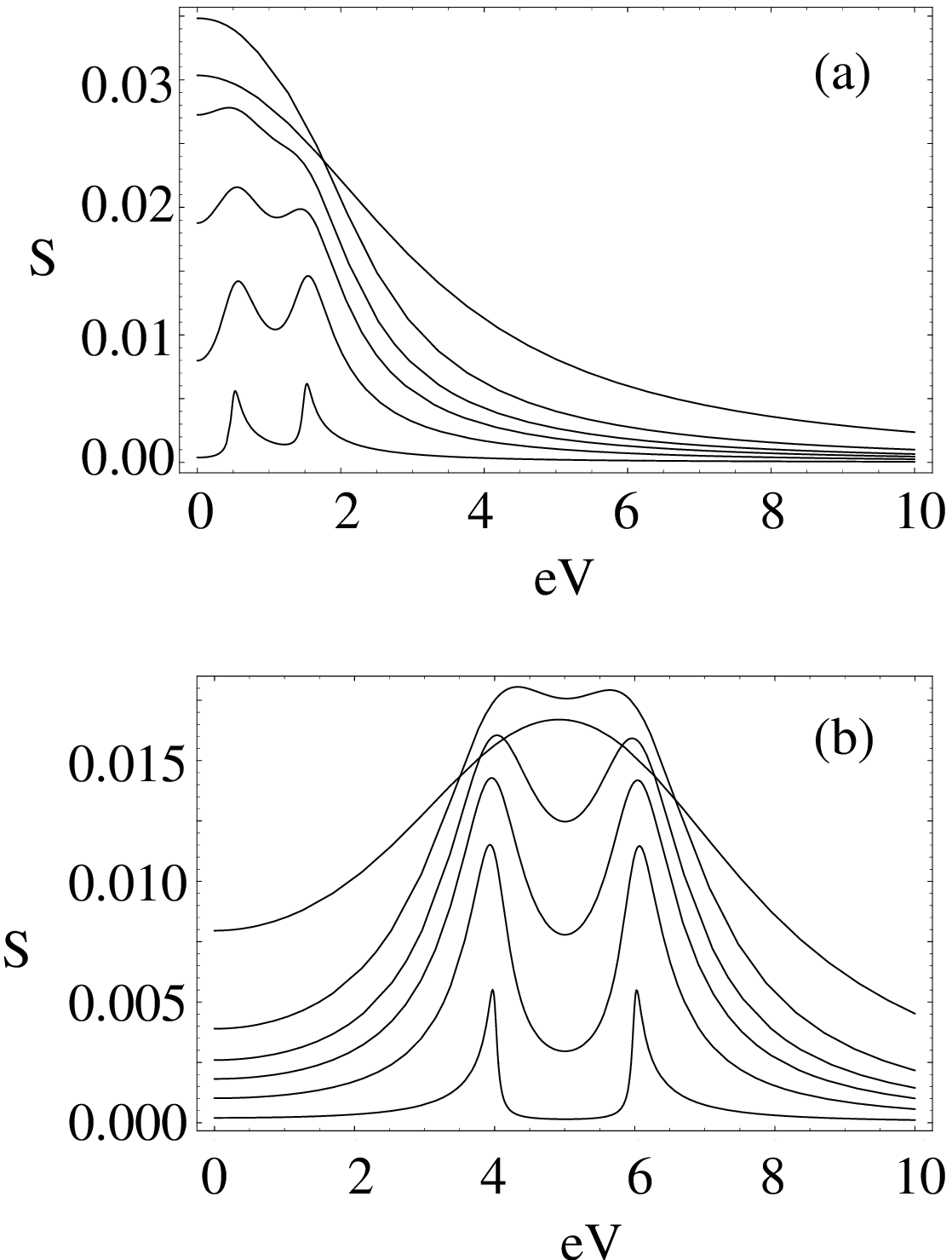}}}
\begin{figure}
\caption{The same as in figure 2 for finite exchange energy, 
(a) $h=0.5\epsilon_T$ and (b) $h=5\epsilon_T$.}
\label{fig4}
\end{figure}

{\centerline{\epsfxsize=\figwidth\epsfbox{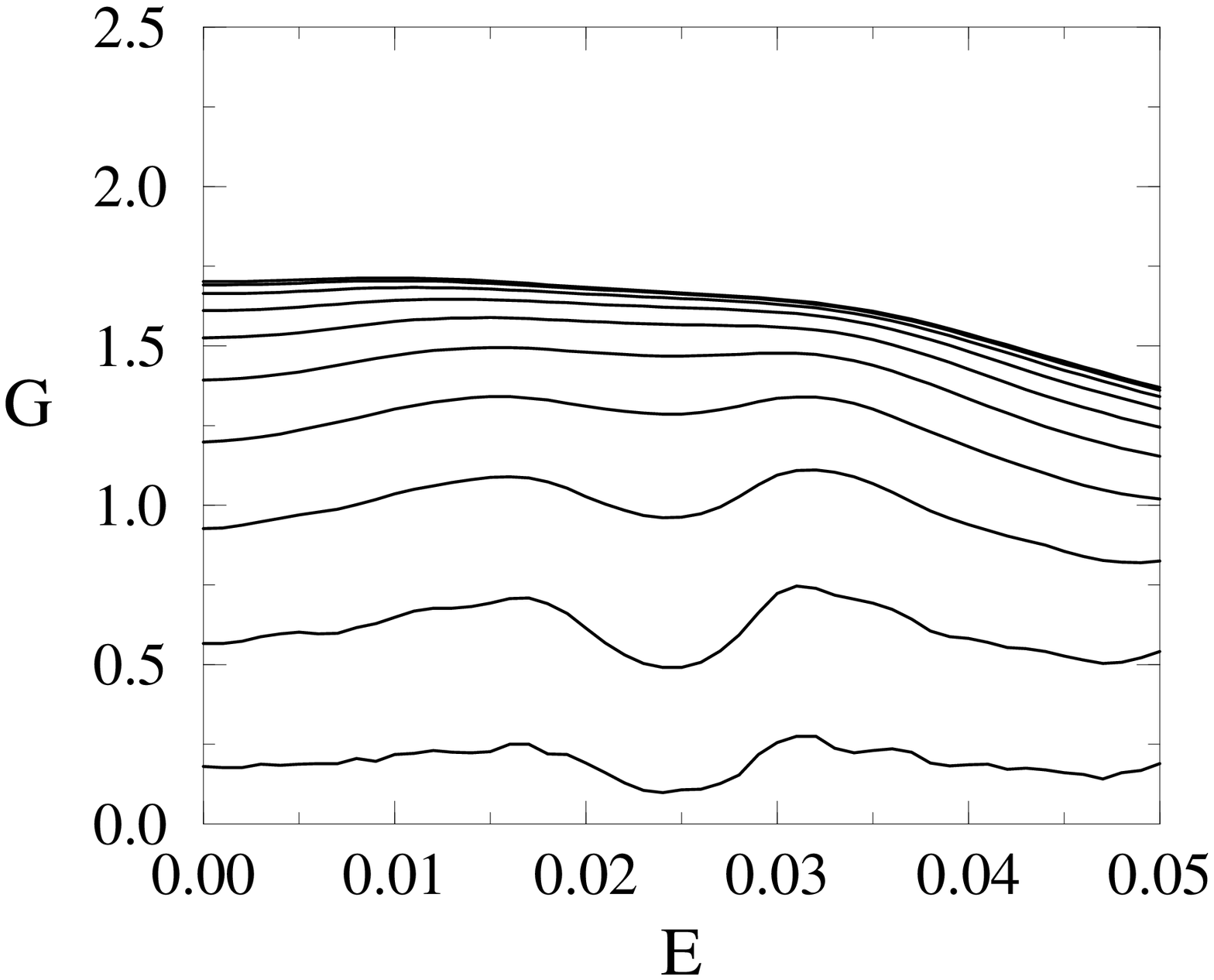}}}
\begin{figure}
\caption{The same as in figure 3 for finite exchange energy, 
$h=0.025$.}
\label{fig5}
\end{figure}

{\centerline{\epsfxsize=\figwidth\epsfbox{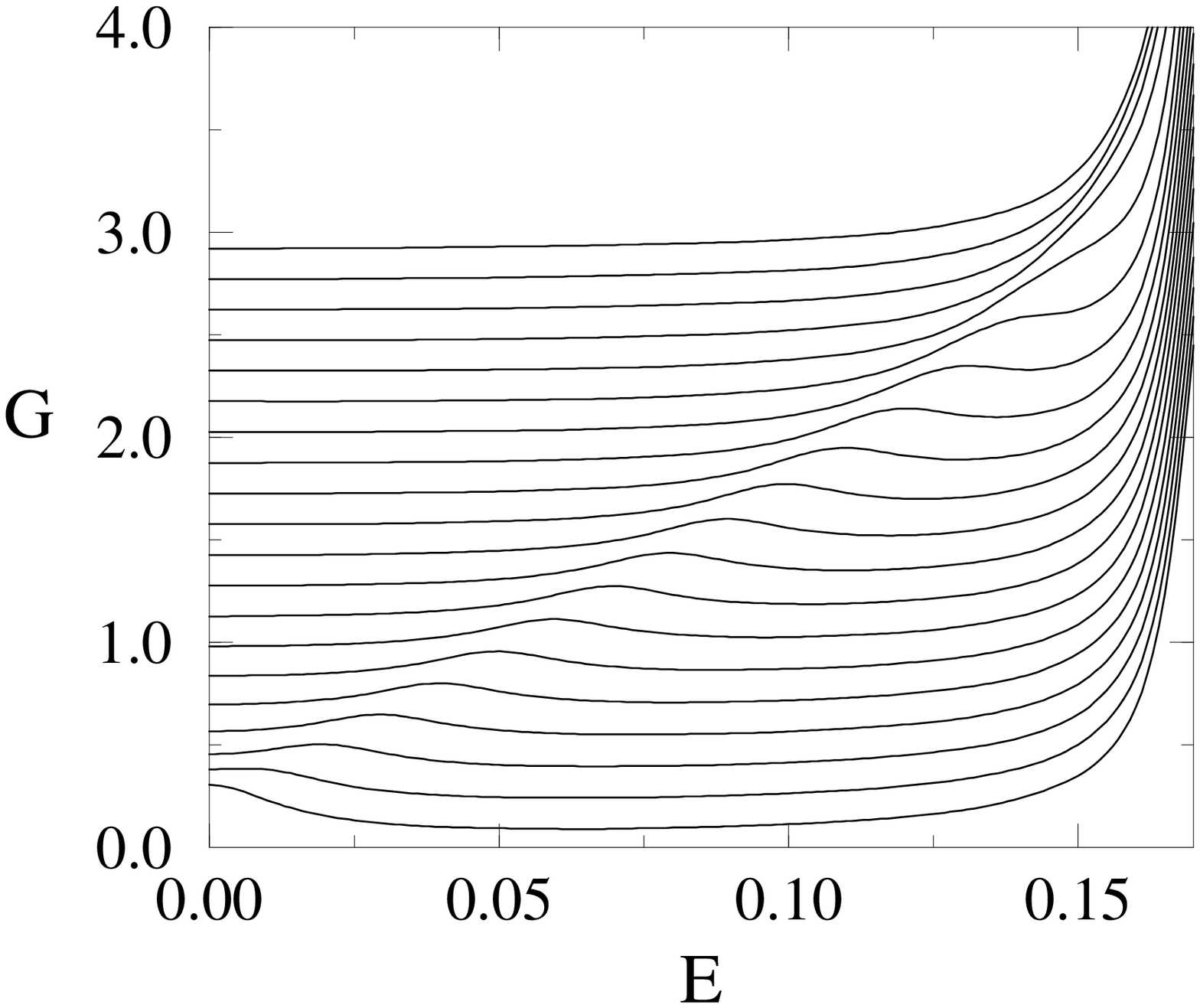}}}
{\centerline{\epsfxsize=\figwidth\epsfbox{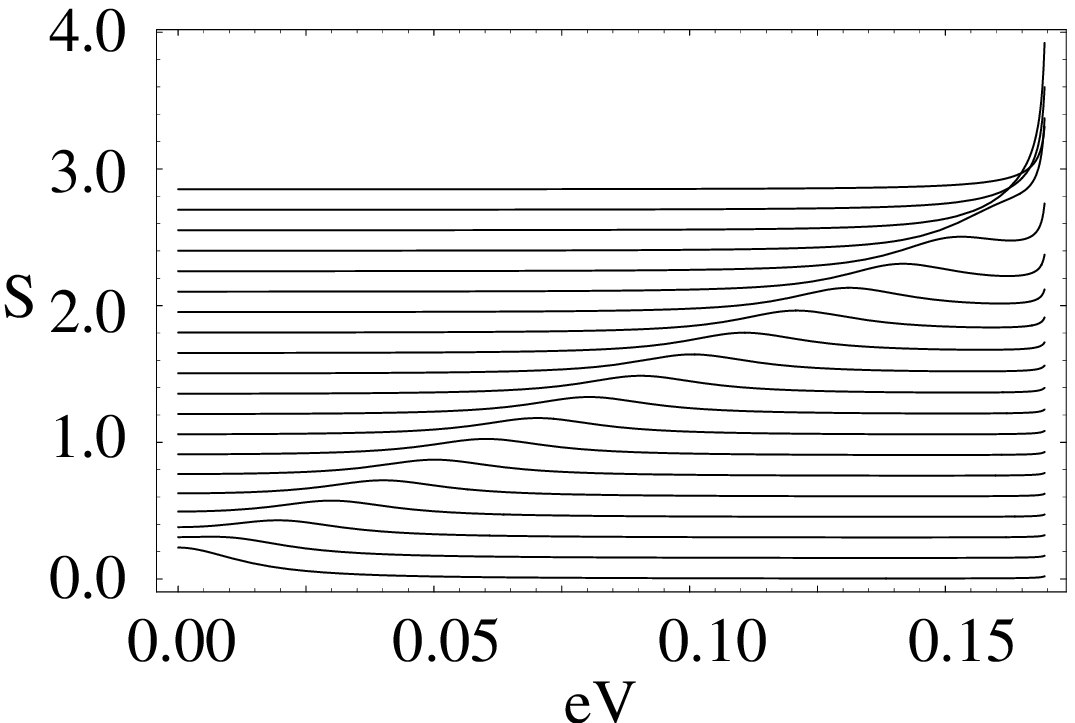}}}
\begin{figure}
\caption{(a) The conductance $G$ vs voltage as
obtained from numerical simulations,  for 
varying  exchange energy, $h$,  from $h =0$ (lowermost curve) to $h
=0.19$ (uppermost curve). These results are in the regime 
$\gamma\gg\epsilon_T$. Each
subsequent curve was translated vertically by an amount $0.15$.
A value of $\Delta=0.17$ was used.
(b) The normalised conductance $S$ vs voltage as
obtained from quasiclassical theory,  for 
varying exchange energy, $h$,  from $h =0$ (lowermost 
curve) to $h =0.19$ (uppermost curve). Each curve is shifted by
0.15 vertically. Values for the zero
voltage conductance and gamma were obtained by fitting to the
numerical result for $h=0$ and the same value of $\Delta$ was used. 
}
\label{fig6}
\end{figure}

\end{multicols}

\end{document}